\documentclass[english]{article}
\usepackage[latin9]{inputenc}
\usepackage{amssymb}
\usepackage{amsmath}
\usepackage{graphicx}
\usepackage{hyperref}
\usepackage{comment}
\hypersetup{
    colorlinks=true,
    linkcolor=blue,
    filecolor=magenta,      
    urlcolor=cyan,
}

\usepackage[]{qcircuit}
\usepackage{braket}
\usepackage[title]{appendix}
\usepackage[linesnumbered,ruled,vlined,algosection]{algorithm2e}
\usepackage{ bbold }
\usepackage[paperwidth=21cm,paperheight=29.7cm]{geometry}
\usepackage{amsthm}
\usepackage{babel}
\usepackage{authblk}
\usepackage{float}
\usepackage{xcolor}
\usepackage{pgfplots}
\usepackage[hang,flushmargin]{footmisc}
\pgfplotsset{compat=1.17}
\newcommand{\red}[1]{{\color{red} #1}}
\newcommand{\blue}[1]{{\color{blue} #1}}

\newcommand{\ddj}{Deutsch-Jozsa }

\newtheorem{theorem}{Theorem}[section]
\newtheorem{remark}[theorem]{Remark}
\newtheorem{example}[theorem]{Example}
\numberwithin{equation}{section}
\title{Quantum advantage and noise reduction in distributed quantum computing}
\author{J. Avron, Ofer Casper and Ilan Rozen }
\affil{Department of Physics, Technion, 320000 Haifa, Israel}

\begin{document}

\maketitle
\begin{abstract}
Distributed quantum computing can give substantial noise reduction due to shallower circuits. An experiment illustrates the advantages in the case of  Grover search.  This motivates studying the quantum advantage of the distributed version of the Simon and \ddj  algorithm. We show that the distributed Simon algorithm retains the exponential advantage, but the complexity deteriorates from $O(n)$ to $O(n^2)$, where $ n = \log_2(N)$. The distributed \ddj deteriorates to being probabilistic but retains a quantum advantage over classical random sampling.
\end{abstract}

\section{Introduction}

The \href{https://quantum-computing.ibm.com/}{IBM quantum experience} and \href{https://aws.amazon.com/braket/}{Amazon Braket} offer the opportunity to implement quantum algorithms on many small and noisy quantum computers.  
{More than 20 quantum computers, with at most 65 qubits} have been deployed by IBM. None can communicate quantumly.
The question then begs itself what  advantages and disadvantages distributed quantum computing with classical communication offer.

Replacing quantum by classical resources usually leads to large overhead. For example, simulating $n$ qubits  needs $O(N=2^n) $ classical bits. More generally, simulating a quantum circuit with $n+k$ qubits by a quantum circuits with $n$ qubits requires\footnote { $c$ is a function of $n$.} $O(2^{ c k})$ uses of the quantum circuits \cite{bravyi2016trading}. 

How much quantum advantage survives in distributed computing depends on the algorithms. Cirac et.~al.~\cite{distributedCIRAC} showed that distributed 3SAT  retains quantum advantage. Bravyi et.~al~\cite{bravyi2016trading} estiumated the overhead in classical computation for sparse quantum circuits and  Peng et.~al.~\cite{peng2020simulating} derived related results for tensor networks with limited connections between clusters.

Distributed quantum computing can offer, besides the obvious advantage of additional ``virtual qubits'',  the advantage of significant noise reduction. This comes about because splitting an algorithm can result in  significant reduction of depth. Since the noise in the output scales exponentially with the depth of the circuit this can be a significant advantage. For example, if the depth of a circuit is large enough the output of the quantum computer may be overwhelmed by noise, but a distributed computation with shallower depth   may  give significant results. As far as we an tell, this simple, one may say trivial, point has not been studied before.

We shall describe an experiment involving Grover search\cite{GroverAlg} that illustrate the advantage of distributed quantum computing over the undistributed computation.   

In addition we study the quantum advantage of distributed algorithms, for two basic textbook examples: Simon's \cite{simon,shor} and \ddj \cite{deutsch}. Since the Grover and \ddj algorithm involve the use of an Oracle we shall also consider the task of distributing an Oracle. 

As we shall see:
\begin{itemize}
\item The distributed Simon's algorithm retains the exponential speedup albeit with higher complexity, see section \ref{s:s}.
        \item { The  quantum advantage of the distributed \ddj deteriorates dramatically, see section \ref{s:dj}.} 
 \end{itemize}

\section{Distributed computations {of Boolean functions}}
{
We restrict ourselves to a particular scheme of distributed computing which suffices to cover the problems we consider. }
\subsection{Distributed classical computation}

{Consider a (classical) circuit with $n$  bits that computes the function}
\begin{equation}
    f:\{0,1\}^n\mapsto \{0,1\}^{m}, \quad m\le n-1
\end{equation}
{The function can be split into its even and odd parts: $f_{even/odd}:\{0,1\}^{n-1}\mapsto \{0,1\}^{m}$ } 
\begin{equation}\label{e:eo}
    f_{even}(y_1,\dots,y_m)= f(y_1,\dots,y_m,0), \quad f_{odd}(y_1,\dots,y_m)= f(y_1,\dots,y_m,1)
\end{equation}
More generally, 
\begin{align}\label{e:p}
    f_{even}(y_1,\dots,y_m)&= f(y_1,\dots,y_{j-1},0,y_{j+1},\dots,y_{m+1}),\nonumber \\ f_{odd}(y_1,\dots,y_m)&= f(y_1,\dots,y_{j-1},1,y_{j+1},\dots,y_{m+1})
\end{align}
{We assume that the even and odd parts can be computed by a circuit with $n-1$ bits. We can then distribute computing $f$ to two $n-1$ bits processors. }

\subsection{Distributed quantum computation}

{Suppose Alice has a processor with $n$ connected qubits and Bob has two devices with $n-1$ connected qubits each. A general state of Alice's processor}
\begin{equation}
    \ket{\psi}=\sum_{j=0}^{N-1} \psi_j \ket j, \quad N=2^n
\end{equation}
{is described by $N$ amplitudes. The state of Bob's two processors is}
\begin{equation}
    \ket{\phi_1}\otimes\ket{\phi_2}, \quad \ket{\phi_j}=\sum_{k=0}^{N/2-1} \phi_{jk}\ket{k}
\end{equation}
{and it, too, is described by $2N/2=N$ amplitudes.   }

Alice's {quantum advantage come from her ability to entangle her $n$ qubits. Bob can not entangle the two processors and can only entangle $n-1$ qubits on each processor. Bob has more qubits and shallower circuits. This is an advantage in noisy quantum computers. Bob also has the advantage that his measurements give  $2(n-1)$ bits of information while  Alice's measurement gives her only $n$  bits of information. We shall, however, not make use of this advantage here. }  

\subsection{Quantum circuits and  DNF }\label{s:dnf}

{ Grover and \ddj algorithms involve the use of Oracles. 
We therefore face the problem of  distributing an Oracle without introducing bias. To do  so we assume that the Oracle is an algorithm with a standard form\footnote{In computer science one normally does not worry how the Oracles does what it does. But, for  the case at hand, we need to. } that allows to split it to the even and odd arguments.}
{ The the reader who is willing to accept this on good faith  may want to skip to the next section.

The Oracle, by definition, computes a Boolean function. Any}  Boolean function can be represented by  DNF. (For an elementary introduction see appendix \ref{a:dnf}). For example,  the DNF of the function that assigns $1$ to $101$ and $010$ and $0$ otherwise is
\begin{equation}\label{e:fstan}
    f(x_0,x_1,x_2)= x_0\cdot \bar x_1\cdot x_2 +\bar x_0 \cdot x_1\cdot \bar x_2
\end{equation}
where $x_j\in \{0,1\}$ are logical variables (equivalently,  binaries) and $\bar x_j$ is the (logical) NOT, (equivalently, for binary $\bar x_j=x_j\oplus 1$). The $+$ is the (logical) OR  normally denoted $\lor$. 

In the general case with $n$ logical variables $x_j, \ j\in 1,\dots,n$, the DNF could be  a sum of a large number of terms, each term is a product over all logical variables; Each  variable appears once either as $x_j$ or as $\bar x_j$.

\begin{figure}[ht]
    \centering
\[
\raisebox{-20pt}{$U_{f} :$}\quad 
\Qcircuit @C=1em @R=1em { 
	& \qw&\qw & \ctrl{2} & \qw & \qw & \gate{X} & \ctrl{2} & \gate{X} & \qw \\
	&\qw& \gate{X} & \ctrl{1} & \gate{X} & \qw & \push{\rule{0em}{1.2em}} \qw & \ctrl{1} & \qw & \qw \\
	& \qw&\push{\rule{0em}{1.2em}} \qw & \ctrl{-2} & \qw &\qw & \gate{X} & \ctrl{-2} & \gate{X} & \qw
	}
	\]
    \caption{The  circuit corresponding to the DNF in Eq.~(\ref {e:fstan}). The columns of control gates is $C^{\otimes n}Z$ gate, in  notation that manifests the symmetry of the gate.}
    \label{f:standard}
\end{figure}
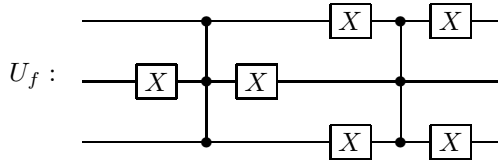

The quantum circuit that computes
\begin{equation}
    U_f\ket{x}= (-1)^{f(x)}\ket{x}
\end{equation}
can be read out from the DNF.  For the DNF in Eq.~(\ref{e:fstan})
the circuit is given in Fig. \ref{f:standard}.
In the general cased,  a pair of $X$ gates decorate  all the $\bar x$ and the n-fold product is represented by $C^nZ$.

\begin{remark}
The DNF is, in general, not the most compact representation of the function $f$  and similarly, the corresponding quantum circuit need not be the optimal circuit. Optimization of quantum circuits is considered in \cite{Bae2020}.
\end{remark}

The even and odd parts of $f$ are easily constructed from the DNF. If we use $x_0$ as the bit that determine even/odd then $f_{e/o}$ for the example in Eq.~(\ref{e:fstan}) is
\begin{equation}\label{e:dnfe}
    f_e=  x_1\cdot \bar x_2, \quad f_o=\bar x_1\cdot x_2
\end{equation}
In the general case, all the  $\bar x_0$ terms make the even part (with $\bar x_0$ deleted) all the  $ x_0$ terms make the odd part (with $ x_0$ deleted).
The DNF is then used to construct the $n-1$ qubits circuits for the even and odd parts:

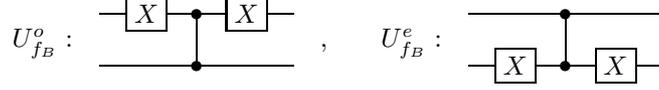
\begin{figure}[ht]
    \centering
\begin{equation*}\label{e:ex2}
\raisebox{-10pt}{$U_{f_B}^o:$}
\quad\Qcircuit @C=1em @R=1.2em { 
	& \gate{X} & \ctrl{1} & \gate{X} &  \qw \\
	& \qw & \ctrl{-1}  & \qw  & \qw
	}\quad\raisebox{-10pt}{,}
\qquad 
\raisebox{-10pt}{$U_{f_B}^e:$}
\quad
\Qcircuit @C=1em @R=1.2em { 
     &  \qw & \ctrl{1} & \qw & \qw \\
	 &  \gate{X} & \ctrl{-1} & \gate{X} & \qw
	 }
\end{equation*}
    \caption{The circuits corresponding to the DNF of Eq.~(\ref{e:dnfe})}
    \label{f:dnfe}
\end{figure}

In conclusion, given Alice's $n$ qubits  circuit corresponding to the DNF, there is a simple procedure that constructs the $n-1$ qubits circuits of Bob for the even and odd parts of the function.

An algorithm for generating the even and odd circuits is:

\begin{algorithm}[H] 
\KwResult{two (n-1)-qubit circuits $ U_{f_B}^{e}, U_{f_B}^{o} $ } 
\SetKwProg{Divide}{Divide}{}{} 
\Divide{$(U_{f_A})$}{
$ parity \gets 0 $\;
$ U_{f_B}^{o} \gets \mathbb{1}$\; 
$ U_{f_B}^{e} \gets \mathbb{1}$\;
\If{$U_{f_A}$ is not of DNF}{abort \;}
\ForEach{ gate $G$ in $U_{f_A}$ in the order they are executed from top to bottom}
{ 
    \uIf{ $G$ is an $X$ gate acting on the parity qubit \footnote{The parity qubit is defined as the qubit that distinguishes between the even and odd subspaces}}{
        $parity \gets NOT(parity) $ \;
    }
    \uElseIf{ $G$ is a $C^{ \otimes (n-1)}Z $}{
        \uIf{ $parity == 1$} {
            append $C^{\otimes(n-2)}Z$ to $U_{f_B}^{e}$\;
        }
        \Else{
            append $C^{\otimes(n-2)}Z$ to $U_{f_B}^{o}$\;
        }
    }
    \uElseIf{ $G$ is a single qubit gate acting not on the parity qubit} {
        append $G$ to its respective qubit both in $U_{f_B}^{o}$ and $U_{f_B}^{e}$ \;
    }
    \Else{
        abort\;
    }
}
return $ U_{f_B}^{o}, U_{f_B}^{e} $ \;
} 
\caption{Splitting a Circuit} 
\end{algorithm}
\vspace{0.3cm}
The algorithm returns the requisite $U_{f_B}^{o}$ and $U_{f_B}^{e}$:
\begin{equation}
    U_{f_A}\ket{1}\otimes\ket{\varphi} \equiv \ket{1}\otimes U^{o}_{f_B}\ket{\varphi} \quad  U_{f_A}\ket{0}\otimes\ket{\varphi} \equiv \ket{0}\otimes U^{e}_{f_B}\ket{\varphi}
\end{equation}

In appendix B we shall describe the converse procedure whereby starting from Bob's distributed quantum circuits, Alice can generate her single quantum circuit.


\section{The depth of distributed computation} \label{sec:shallow}

Distributed algorithms have an advantage of shallower depth.
This is evident for  circuits given by DNF:   The depth of the $n$ qubits circuit is distributed between the two $n-1$ circuits. (See Eqs.~(\ref{e:fstan},\ref{e:dnfe}).)

In the experiment, section \ref{s:ex}, the depth of the distributed circuits was about a factor 4 smaller than the  undistributed circuit. This is a consequence of the fact that the depth of a  $C^nZ$ circuit is considerably larger than the depth of $C^{n-1}Z$. We illustrate this with an example.

{The gain in depth depends on the choice of gates.} We use the following rules  \cite{elementaryGates}:
\begin{itemize}
    \item Any single qubit  gate is allowed
    \item The only 2-qubit gate allowed is CNOT
    \item Gates that can be executed in parallel are grouped into columns
    \item Consecutive single qubits gates are counted as a single qubit gate
\end{itemize}

Consider the circuit in Fig.~\ref{f:standard} and its distributed cousins Fig.~\ref{f:dnfe}. 
To compute $\text{depth}\big(U_{f_A}\big)$ and $\text{depth}\big(U_{f_B}\big)$. 
We first observe that 
\begin{equation}
\text{depth}(C^2Z) = 11
\end{equation}
This is seen from Fig. \ref{f:c2z}
\begin{figure}[h]
    \centering
\begin{equation*}
\Qcircuit @C=1em @R=1.9em { 
	& \ctrl{2} & \qw & & & & \\
	& \ctrl{1} & \qw & & \equiv & & \\
	& \ctrl{-2} & \qw & & & &
}
\Qcircuit @C=1em @R=.9em { 
    & \qw  & \qw & \ctrl{2} & \qw & \qw & \qw & \ctrl{2} & \qw  & \ctrl{1} & \gate{T} & \ctrl{1} & \qw \\
    & \ctrl{1} & \qw & \qw & \qw & \ctrl{1} & \qw & \qw & \gate{T} & \targ & \gate{T^\dagger} & \targ & \qw \\
     & \targ & \gate{T^\dagger} & \targ & \gate{T} & \targ & \gate{T^\dagger} & \targ & \gate{T} & \qw & \qw & \qw & \qw
}
\end{equation*}

    \caption{$C^2Z$}\label{f:c2z}
    \label{fig:my_label}
\end{figure}

It follows that 
\begin{equation}
    \text{depth}(U_{f_A})=3 + 2 \cdot \text{depth}(C^2Z) = 25
\end{equation}
To compute the depth of the corresponding distributed circuits in Fig.~\ref{f:dnfe} we first recall
\begin{equation}
\Qcircuit @C=1em @R=1.2em { 
	& \ctrl{1} & \qw & & \equiv & & \\
	& \ctrl{-1} & \qw & & & &
}
\Qcircuit @C=1em @R=1.2em { 
	& \qw & \ctrl{1} & \qw & \qw \\
	& \gate{H} & \targ & \gate{H} & \qw
}
\end{equation}
For the even circuit one   combines the $X$ gates with the Hadamard gates. Hence both circuits have:
\begin{equation*}
    \text{depth}(U_{f_B}^{e/o}) = 3
\end{equation*}
The example shows that distributed circuits can lead to substantial reduction in the depth of  quantum circuits.

\section{Distributed Grover: An experiment }\label{s:ex}

In this section  we describe an experiment carried on the IBMQ5 (ibmq\_santiago). The data for the machine (at the time of the experiments) are given in Table \ref{t:ibm}.
\begin{table}[h]
\begin{center}
    \begin{tabular}{|l|c|}
    \hline
    Single qubit error    &$ O( 2.2\times 10^{-4})$  \\
    Two qubit   error   & $\epsilon\approx 6.2\times 10^{-3}$\\
    Coherence & $T_c\approx 133\ [\mu~ sec]$\\
    Gate rate & $R\approx 408\  [\eta~sec]$\\
    \hline
    \end{tabular}
    \caption{Machine data for IBM santiago}\label{t:ibm}
\end{center}
\end{table} 

In the experiment a Grover search for a single target among 16 items, ($N=16$ and $M=1$) was conducted on an undistributed circuit with $n=4$ qubits and then on  distributed circuits with $n=3$ qubits each. The target state in both cases has been $\ket{1111}$. 

{We made use of the open-source \href{https://qiskit.org/}{Qiskit} library to generate the Grover search circuits which were then run on real (i.e not simulated) machines}. The circuits data are given in Table \ref{t:3-4}.

\begin{table}[h]
\begin{center}
    \begin{tabular}{|l|c|c|c|c|c|c|}
    \hline
    Qubits    & Gates & CNOT& Time steps&Iterations& Repetitions &Success probability \\ 
    \hline
    4   & 519&291& 396& 3&8096& 0.063\\
    3 (odd) & 140 &55&93& 2&8096& 0.4\\
    3 (even) & 59&24&39& 2& 8096& NA\\
    \hline
    \end{tabular}
    \caption{Circuit data for the undistributed computation with $n=4$ qubits and the two distributed circuits with $n=3$ qubits. The numbers in the table are for the transpiled circuits.}\label{t:3-4}
\end{center}
\end{table}

Fig.~\ref{f:G4}, shows the results for the undistributed Grover and Figs.~\ref{f:g3} for the distributed Grover search.    Clearly, the undistributed  search  failed, while  the distributed search  qubits succeeded in finding $\ket{1111}$.

{The results are in agreement with what one should expect for noisy machines that can handle limited depth. For more details about the expected and observed noise in the circuits, see Appendix \ref{ap:error}. }

\subsection{Undistributed Grover search}

\begin{center}
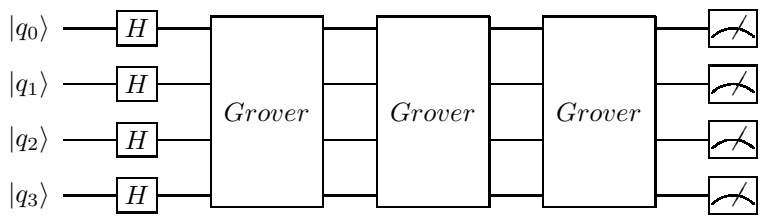
\begin{figure}[h!]
    \centerline{
    \Qcircuit @C=1em @R=.7em 
    { 
	    & \lstick{\ket{q_0}} & \qw & \gate{H} & \qw & \multigate{3}{Grover} & \qw & \multigate{3}{Grover}  & \qw &  \multigate{3}{Grover} & \qw 
	    & \meter \\
	& \lstick{\ket{q_1}} & \qw & \gate{H} & \qw & \ghost{Grover} & \qw & \ghost{Grover} & \qw & \ghost{Grover} & \qw &
	\meter \\
	& \lstick{\ket{q_2}} & \qw & \gate{H} & \qw & \ghost{Grover} & \qw & \ghost{Grover} & \qw &  \ghost{Grover} & \qw &	\meter \\
	& \lstick{\ket{q_3}} & \qw & \gate{H} & \qw & \ghost{Grover} & \qw & \ghost{Grover} & \qw &  \ghost{Grover} & \qw 
	& \meter
    }
    }
    \caption{The undistributed Grover search circuit with $n=4$ qubits.
    }
    \label{f:circ-Grover-4}
\end{figure}
\end{center}
The optimal number of Grover iterations  with $n=4$ is {$r=(\pi/4)\sqrt 16=\pi\approx 3$} and the {(theoretical)} success probability is\footnote{$p=\sin^2((2r+1) \theta)\approx0.96$ where , $ p=\sin  \theta = 1 /{\sqrt N}$} $0.96$.
The circuit produced the histogram in Fig.~\ref{f:G4} and manifestly  failed to identify the target $\ket{1111}$.

\begin{center}
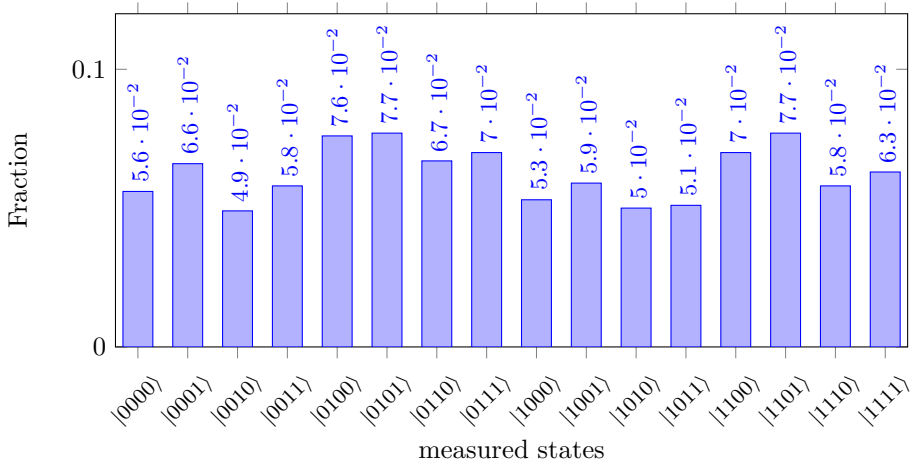
\begin{figure}[h!]
    \centering
    \begin{tikzpicture}
    \begin{axis}[
        width=12cm,
        height=6cm,
        ybar,
        bar width=0.6,
        ymin=0,
        ymax=0.12,
        ytick={0,0.1},
        xticklabel style={rotate=45,font=\footnotesize},
        xticklabels from table={hist_data/bighist.dat}{x},
        xtick=data,
        enlarge x limits=0.03,
        xlabel=measured states,
        ylabel=Fraction,
        x label style={at={(axis description cs:0.5,-0.25)},anchor=north},
        y label style={at={(axis description cs:-0.1,.5)},anchor=south},
        every node near coord/.append style={rotate=90,font=\small},
      nodes near coords align={horizontal},
        nodes near coords
    ]
    \addplot table[x expr=\coordindex,y=y] {hist_data/bighist.dat};
    \end{axis}
    \end{tikzpicture}
    \caption{{The histogram for the undistributed Grover algorithm with  $n=4$ qubits searching for the target state $\ket{1111}$. The sampling error is $ \pm 0.011 $.  The  search failed}.}
    \label{f:G4}
\end{figure}
\par\end{center}

It is useful to extract form the histogram the frequencies of bit-flip error in the target $\ket{1111}$. This is shown in Table \ref{t:f4}. The data agree with the Binomial distribution up to $\pm \sigma$. The large error probability for bit-flip, $p_0\approx0.50$, is the reason for the failure of the search. $p_0$ is in reasonable agreement with the estimates based on the machine data in Table \ref{t:ibm}, see Appendix \ref{ap:error}. 

\begin{table}[h]
\begin{center}
\begin{tabular}{|c||c|c|c|c|c|}
\hline
\# errors   &  0&1&2&3&4\\
\hline
Frequency     & 0.06& 0.26&0.38&0.25&0.06\\
\hline
Binomial[$n=4,p_0\approx 0.50$]&0.06&0.25&0.38&0.25&0.06\\
\hline
\end{tabular}
\caption{The bit flip error frequency in the search for the target $\ket{1111}$ is well approximated by the Binomial distribution.}\label{t:f4}
\end{center}
\end{table}

\subsection{Distributed Grover search} 
The distributed Grover search runs with two Oracles, one for the 'odd' subspace and the other
for the 'even' subspace, both with $n=3$ qubits. The target $\ket{1111}$ is encoded in the odd Oracle and  the even Oracle is ``empty". The optimal number of iterations in the odd circuit is {$r=(\pi/4)\sqrt 8\approx 2$} with success probability $p\approx 0.95$. 
The circuits are {shown in Fig.~\ref{fig:my_label4}}. 

The left histogram in Fig.~\ref{f:g3} clearly identifies the correct answer $\ket{111}$ (corresponding to $\ket{1111}$). The even circuit has no marked element\footnote{Although the formal optimum for $M=0$ is $r=\infty$, the optimal number of Grover iterations is actually $r=0$.} {and this is borne out by the histogram on the right.} 
\begin{flushleft}
\begin{figure}[h!]
    \centering
    \centerline{
    \Qcircuit @C=1em @R=.7em 
    { 
	    & \lstick{\ket{q_0}} & \qw & \gate{H} & \qw & \multigate{2}{Grover_{e/o}} & \qw & \multigate{2}{Grover_{e/o}}  & \qw & \meter \\
	    & \lstick{\ket{q_1}} & \qw & \gate{H} & \qw & \ghost{Grover_{e/o}} & \qw & \ghost{Grover_{e/o}} &  \qw & \meter \\
	    & \lstick{\ket{q_2}} & \qw & \gate{H} & \qw & \ghost{Grover_{e/o}} & \qw & \ghost{Grover_{e/o}} & \qw   & \meter \\ \\
    }
    }
    \caption{The distributed circuits with $n=3$  for the even/odd arguments.  }  
    \label{fig:my_label4}
\end{figure}
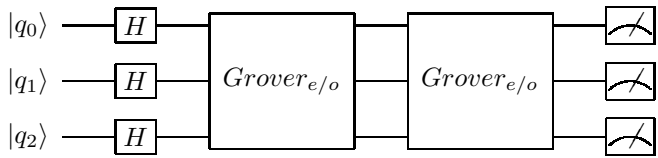
\end{flushleft}

\begin{center}
\begin{figure}[h!] 
    \centerline{
    \begin{tikzpicture}[scale=.8]
    \begin{axis}[
        width=9cm,
        height=7cm,
        ybar,
        bar width=0.35,
        ymin=0,
        ymax=0.5,
        xticklabel style={rotate=45,font=\footnotesize},
        xticklabels from table={hist_data/oddhist.dat}{x},
        xtick=data,
        enlarge x limits=0.03,
        xlabel=measured states in the odd sub-space,
        ylabel=Fraction ,
        x label style={at={(axis description cs:0.5,-0.2)},anchor=north},
        y label style={at={(axis description cs:-.1,.5)},anchor=south},
        minor y tick num = 10,
        every node near coord/.append style={rotate=90,font=\small},
        nodes near coords align={horizontal},
        nodes near coords
    ]
    \addplot table[x expr=\coordindex,y=y] {hist_data/oddhist.dat};
    \end{axis}
    \end{tikzpicture}
    \begin{tikzpicture}[scale=.8]
    \begin{axis}[
        width=9cm,
        height=7cm,
        ybar,
        bar width=0.35,
        ymin=0,
        ymax=0.33,
        xticklabel style={rotate=45,font=\footnotesize},
        xticklabels from table={hist_data/evenhist.dat}{x},
        xtick=data,
        enlarge x limits=0.03,
        xlabel=measured states in the even sub-space,
        x label style={at={(axis description cs:0.5,-0.2)},anchor=north},
        y label style={at={(axis description cs:-0.2,.5)},anchor=south},
        minor y tick num = 10,
        every node near coord/.append style={rotate=90,font=\small},
        nodes near coords align={horizontal},
        nodes near coords
    ]
    \addplot table[x expr=\coordindex,y=y, y index = {1}] {hist_data/evenhist.dat};
    \end{axis}
    \end{tikzpicture}
    }
    \caption{The two histograms for the distributed search with $n=3$ qubits.   The sampling error is $ \pm 0.011 $. The left histogram  correctly identified $\ket{111}$. The right histogram did  not single out any state, as it should. }
    \label{f:g3}
\end{figure}
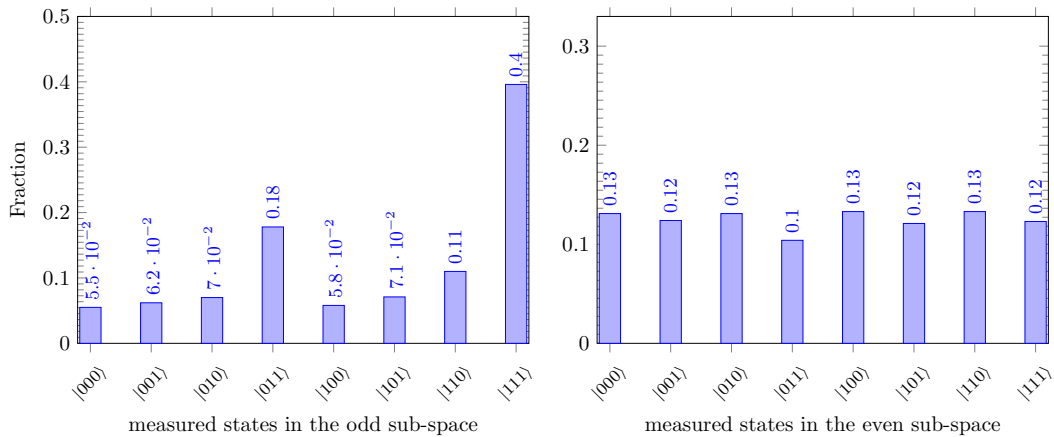
\end{center}
It is instructive to look at the probability of bit flip  error. The histogram in Fig.~\ref{f:g3} leads to Table \ref{t:error3}.   The agreement with a Binomial distribution is only qualitative.  The small bit flip error $p_0\approx 0.30$ is why the search succeeded. $p_0$ can be estimated from the machine data in Table \ref{t:ibm}, see Appendix \ref{ap:error}.
\begin{table}[h]
\begin{center}
\begin{tabular}{|c||c|c|c|c|}
\hline
\# errors   &  0&1&2&3\\
\hline
Frequency     & 0.40& 0.36&0.19&0.05\\
\hline
Binomial[$n=3,p_0\approx 0.30$]&0.35&0.44&0.19&0.03\\
\hline
\end{tabular}
\caption{The bit flip error frequencies in the distributed search. }\label{t:error3}
\end{center}
\end{table}

\section{Distributed Simon's algorithm}
\subsection{Period finding and the~\texorpdfstring{$\mathbb{Z}_2^n$}{Z2n}  Fourier transform}\label{s:s}

   \begin{center}
\begin{figure}[ht]
\centerline{
   \Qcircuit @C=1em @R=1.2em { 
\lstick{\ket{0}^{\otimes n}}	& \gate{H^{\otimes n}} & \gate{(-1)^f}  & \gate{H^{\otimes n}} & \qw &\meter
}
}
    \caption{The \ddj and the period finding circuit }
    \label{fig:ddj}
\end{figure}
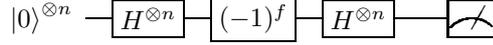
        \end{center}

{Consider   $f:\{0,1\}^n\mapsto\{0,1\}$, a Boolean function periodic with period $s\neq 0$ under bit-wise addition}\begin{equation}
    f(x\oplus s)=f(x) \quad \forall x\in\{0,1\}^n, 
\end{equation}
{Task: Find $s$.} This is a special case of  Simon's problem \cite{simon}.

Period finding is the business of Fourier transforms. For the case at hand, this is the $\mathbb{Z}_2^n$ Fourier transform.
The circuit in Fig. \ref{fig:ddj} reduces period finding to solving a set of linear equations.
Chasing the state $\ket {0}^{\otimes n}$ through the circuit in Fig. \ref{fig:ddj}, using the identity
\begin{equation}
    H^{\otimes n} \ket x=\frac 1 {\sqrt N}\sum_y (-1)^{x\cdot  y} \ket y
\end{equation}
one finds \begin{align}\label{e:out}
    H^{\otimes n}\ket 0&=
    \frac 1 {\sqrt N}\sum_{x}\ket{x} \nonumber\\
    &\xrightarrow{(-)^f}\frac 1 {\sqrt N}\sum_{x} (-1)^{f(x)}\ket{x}  \nonumber \\ &=
    \frac 1 {2\sqrt N}\sum_{x} (-1)^{f(x)}\big(\ket{x}+\ket{x\oplus s}\big) \nonumber \\
    &\xrightarrow{H^{\otimes n}}
    \frac 1 {2 N}\sum_{y} \underbrace{\left(\sum_{x}(-1)^{f(x)+x\cdot y}\right)}_{=g(y)}\big(1 +(-1)^{s\cdot y}\big)\ket{y}
    \end{align}
$g(y)$ is the $\mathbb{Z}_2^n$ Fourier transform of $(-1)^f$, As such it is localized on arguments related to the period as can be seen from
\begin{equation}
    2g(y)=\sum_{x}(-1)^{f(x)+x\cdot y}+\sum_{x}(-1)^{f(x\oplus s)+(x\oplus s)\cdot y}=g(y)\left(1+(-1)^{s\cdot y}\right)
\end{equation}
It follows that:
\begin{equation}
    g(y)=w_y  \delta(s\cdot y)
\end{equation}
$w_y\in\mathbb{Z}$ is the ``weight" of the delta function.
Inserting this to Eq.~(\ref{e:out}) gives for the state exiting the circuit
\begin{equation}\label{e:sim}
    \frac 1 N\sum_{y_j\cdot s =0} w_{y_j}\ket{y_j}
\end{equation}
The outgoing state is therefore a linear combination of solutions of $s\cdot y=0 \mod 2$.
The period is determined as the  solutions of the linear system\begin{equation}\label{e:simon}
    y_j\cdot s=0 \mod 2, \quad w_{y_j} \neq 0
\end{equation}
The random outcomes of the measurement of the quantum circuits give $y_j$.

In the case there are $n$ independent vectors $y_j$,  Eq.~(\ref{e:simon}) has a unique solution $s=0$. In  the case that $n-1$ vectors are independent there is a unique non-trivial period $s$, etc. In the case that the sum in Eq.~(\ref{e:sim}) has a single term $w_0$ for $y=0$, Eq.~(\ref{e:simon}) trivializes and any $s$ is a solution: $f$ is a constant.

\begin{remark}
The weights $\{w_y\}$ are (at most) $N$ integers that satisfy Pythagoras
\begin{equation}\label{e:pyt}
    \sum_{y\cdot s=0} w_y^2=N^2
\end{equation}
This imposes a (Diophantine) constraint on
the allowed $\{w_y\}$ which is independent of the function $f$.
\end{remark}
\begin{example}
With $N=4$, the solutions  $\{w_y\}$ of Eq.~(\ref{e:pyt}) are\footnote{In the sense that for example for the first set $w_y=|4|$ meaning either $w_y=+4$ or $w_y=-4$}
\begin{equation}
      \{ |4|\},\quad \{|2|,|2|,|2|,|2|\}
\end{equation}
Not all of the solutions are realized as Fourier transforms of (the phases of) Boolean functions. In fact, there are 16 Boolean functions of 2 bits: 
\begin{itemize}
    \item 2 constant functions $f(x_1,x_2)=0$ and $f(x_1,x_2)=1$ where $s$ is arbitrary.
    \item 6 balanced functions where $1$ has two pre-images. These have a single non-trivial period.
    \item The 4 functions where $1$ has a single pre-image and 4 where $1$ has three pre-images are a-periodic with $s=0$.
\end{itemize}
The Fourier transforms of the phase functions $(-1)^f$ are
\begin{itemize}
    \item $\pm 4\delta(y)$ for the constant functions.
    \item $\pm 4 \delta(y-j), \ j\in 1,2,3$ for the balanced functions
    \item $\pm 2 \big(1-2\delta(y-j)\big), \quad j\in 0,1,2,3$ for   the a-periodic.
\end{itemize}
\end{example}

Consider the distributed algorithm in the case of a unique $s\neq 0$. Suppose first that $s$ is even (e.g.~$s=10$). The distributed Oracle reduces to the problem for $n-1$ qubits for the even (odd) Oracles. This allows to determine $s$ after $O(n)$ queries.

In the case that $s$ is odd, $x$ and $x\oplus s$ have different parity. The  $n-2$ queries of the distributed algorithm will give the trivial  result $2s=0$. One then needs to try again with a different notion of even-odd, per Eq.~(\ref{e:p}). If the new notion of even-odd gives $s$ even the next $n-2$ queries will determine $s$ after a total of $3(n-2)$ queries. If $s$ is odd, we need to repeat the process. 
The complexity of an algorithm is determined by the worst case corresponding to $n$ repetitions. This gives
\begin{equation}
    O(n^2)
\end{equation}

\begin{remark} 
Similar arguments apply for the standard Simon algorithm for $f:\{0,1\}^n \mapsto \{0,1\}^n$ (which is represented by a quantum circuit acting on $2n$-qubits). Consider two quantum circuits each corresponding to the odd and even sub-spaces, defining functions $f_e,f_o:\{0,1\}^{(n-1)} \mapsto \{0,1\}^n$ (each represented by a quantum circuit acting on $2n-1$-qubits). Following similar steps as in the paragraph above, we find that  complexity in this case is also $O(n^2)$.
\end{remark}

In summary: The complexity of the period finding  of a phase Oracle is:
\begin{itemize}
    \item $O( N)$ classically: The cost of the $\mathbb{Z}_2^n$ Fourier transform.
    \item $O(n)$ for the $n$ qubits quantum circuit
    \item $O(n^2)$ for the distributed quantum circuit.
\end{itemize}

\section{Distributed \ddj algorithm}\label{s:dj}
A Boolean function $f$ is called balanced  if
\begin{equation}
    \sum_x (-)^{f(x)}=\sum_x\big(1-2f(x)\big)=0
\end{equation}
There are 
\begin{equation}
    \binom{N}{N/2}, \quad N=2^n
\end{equation}
 balanced functions. A number which is super-exponentially large. There are, of course, only  two constant Boolean functions: $f \equiv 0$ and $f \equiv 1$.

The \ddj task is: Given the promise that $f$ is either constant or balanced, determine which it is.

If no error is tolerated, one needs $N/2+1$ classical queries of $f$. If one is satisfied with a correct answer with high probability, then few queries suffice. Indeed, the probability that $k$ random queries of a balanced functions have the same image under $f$ is 
\footnote{The formula follows from repeated application of the fact that in an urn with $W$ white stones and $B$ black stone, the probability for picking a black stone is $ B/(B+W)$.
}
\begin{equation}
    p=
    \left(1- \frac N {2N-1} \right)\dots\left(1-\frac N{2N-k+1} \right)
\end{equation}
 When $k\ll N$ one has
\begin{equation}
   p\approx
    \left(\frac 1 2\right)^k  
\end{equation}
If one tolerates $\epsilon$ error probability 
\begin{equation}\label{e:rs}
    k=O(\log 1/\epsilon)
\end{equation}
queries suffice.

The \ddj circuit of Fig.~\ref{fig:ddj} outputs 
\begin{equation}\label{e:dj}
    Prob(y=0)=\left|\frac 1 N \sum_x (-1)^{f(x)}\right|^2=\begin{cases} 1 & f \in const\\
    0 & f \in balanced
    \end{cases}
    \end{equation}
and determines $f$, with no error, with a single query (assuming that the quantum gates are error free and the $f$ is indeed either balanced or constant).  

Now consider the corresponding  distributed \ddj.  The even and odd parts of a constant function are still constant functions. But, the even and odd parts of a balanced function need not be (two) balanced functions. 
The distribution of  $1$ in the even function is the same as randomly drawing $N/2$ stones from two urns with $N/2$ stones each, all $1$ or all  $0$.  
The probability distribution for finding $k$ $1$'s in the even sequence is 
\begin{equation}
    Prob(k)=\binom{N/2}{k} \binom{N/2}{N/2-k}\Big / \binom{N}{N/2}
\end{equation}
We are interested in the probability that the distributed circuit will (mis) identify a balanced function as constant. By Eq.~\ref{e:dj} this is related to the expectation of
\begin{equation}
\left|\frac 1 N \sum_x (-1)^{f_e(x)}\right|^2=\left(1-\frac 2 N \sum_x f_e(x)\right)^2= 1- \frac 4 N \sum_x f_e(x)+ 4\left(\frac 1 N \sum f_e(x)\right)^2
\end{equation}
Evidently
\begin{equation}
    \mathbb{E}\left(\sum_x f_e(x)\right)=\sum_k k \, Prob(k)= \frac N 4
\end{equation}
As $f_e(x)$ and $f_e(y)$ are independent for $x\neq y$ and $f_e^2(x)=f_e(x)$ we also have
\begin{equation}
    \mathbb{E}\left(\sum_x f_e(x)\right)^2=
    \mathbb{E}\left(\sum_{x,y} f_e(x)f_e(y)\right)=
     \mathbb{E}\left(\sum_{x} f_e(x)\right)= \frac N 4
\end{equation}
It follows that the expectation values that a single query of the distributed circuit will make the mistake of identifying a balanced $f$ as constant is 
\begin{equation}
    {\mathbb{E}(y=0| f=\text{balanced})= \frac 1 N}
\end{equation}
Comparing with Eq.~(\ref{e:rs}) we see that one needs  $O(n)$ classical queries to get the same margin of error as a single quantum query.

\bigskip

In summary: The complexity of the \ddj problem is
\begin{itemize}
    \item $1+N/2$ classical queries for a deterministic result.
    \item A single quantum query (for an  error-free circuit).     \item A single query of a distributed  quantum circuits  for  $O(1/N)$ error.
    \item $O(n)$ classical queries for $O(1/N)$  error. 
\end{itemize}

The first two entries imply an $O(N)$ quantum advantage and the last two entries a $O(n)$ advantage of distributed quantum computing.

\section{Conclusion}

Distributed quantum computing with classical communication on ideal devices is, in general, inferior to undistributed quantum computing. However, in the context of the currently available noisy small computers that offer no error corrections,  
distributed computing  offers two advantages: Amalgamating qubits resources and shallow circuits with significant noise reduction. 

\section*{Acknowledgment} We thank Eyal Bairey, Shay HaCohen-Gourgy, Oded Kenneth and Netanel Lindner for helpful discussion and  E.B. for pointing out \cite{Cirac}. We thank the referees for pointing out \cite{bravyi2016trading,peng2020simulating}.

\begin{appendices}

\appendixtitleon

\section{The DNF of quantum circuits}\label{a:dnf}

For the sake of simplicity and concreteness we illustrate an  algorithms that works for any  Boolean function $ f : \{0,1\}^n \rightarrow \{0,1\}^m $ by considering the special case $n=2$ and $m=1$.

\subsection{From  the truth table to DNF  \texorpdfstring{\cite{Hilbert}}{hilbert}}

The Disjunctive Normal Form \cite{DNF} of a Boolean function can be calculated directly from the truth table of the function. Consider:

\begin{displaymath}
\begin{array}{|c c|c|}
x_0 & x_1 & f(x)\\ 
\hline
0 & 0 & 1\\
0 & 1 & 0\\
1 & 0 & 1\\
1 & 1 & 0\\
\end{array}
\end{displaymath}
The DNF form of $f$ is given by
\begin{equation}\label{e:f}
f_{DNF}(x) = \bar x_0 \cdot \bar x_1 + x_0 \cdot \bar x_1
\end{equation}
where $\bar x$ denotes $NOT(x)$. 

We only need to consider the rows where $f(x)=1$. 
For every such row  we  build out of the arguments $x_0,x_1$ a a Boolean statement made of only NOT and AND operations (conjunctive) so that an argument that takes the value $ 1 $ is written as is, while argument with value $ 0 $ is negated. For the  example:

\begin{itemize}
    \item The first row is described by $\bar x_0 \cdot \bar x_1$. 
    \item The third row is described by $ x_0 \cdot \bar x_1 $.
\end{itemize} 
\noindent\fbox{
\parbox{\textwidth}{\underline{Generalization}: In the case of general $ n $ do the same with $n$ variables.  For a different $ m $ repeat the process for each $ f_i(x) $. }
}

\bigskip

The DNF extracted from the truth table is not, in general, the simplest DNF formula of the function. For example $f$ of Eq.~(\ref{e:f}) can be written more simply as:
\begin{displaymath}
f_{DNF} = \bar x_1 
\end{displaymath}
The optimal DNF formula can be calculated using Karnaugh maps \cite{Karnaugh}.

\subsection{From  DNF to  quantum circuit  \texorpdfstring{\cite{BuildingQuantum}}{bogdan}}
\begin{figure}
    \centering
\centerline{
\Qcircuit @C=1em @R=1em {
\lstick{\ket{x_0}} & \gate{X} & \qw & \ctrl{2} & \qw & \gate{X} & \qw & \qw & \qw & \ctrl{2} & \qw & \qw & \qw & \rstick{\ket{x_0}} \\
\lstick{\ket{x_1}} & \gate{X} & \qw & \ctrl{1} & \qw & \gate{X} & \qw & \gate{X} & \qw & \ctrl{1} & \qw & \gate{X} & \qw & \rstick{\ket{x_1}} \\
\lstick{\ket{0}} & \qw & \qw & \targ & \qw & \ustick{\ket{\bar{x}_0 \cdot \bar{x}_1}} \qw & \qw & \qw & \qw & \targ &\qw & \qw & \qw & \rstick{\ket{\bar{x}_0 \cdot \bar{x}_1 + x_0 \cdot \bar{x}_1} 
} 
}
}

    \caption{The DNF circuit for the function gate of Eq.~(\ref{e:f})}
    \label{fig:dc}
\end{figure}
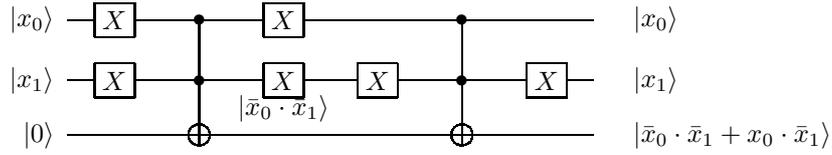

The quantum circuit for any $ f : \{0,1\}^n \rightarrow \{0,1\}^m $ can be built with $X$ gates and $ C^{\otimes n}X $ gates. 
To see this \blue{recall that} Toffoli gate \cite{toffoliLogic} gives the conjunction of its arguments $f_T(x_0,x_1)= x_0\cdot x_1$. 
The quantum circuit for calculating the function in Eq.~(\ref{e:f}) is shown in Fig.~\ref{fig:dc}. 
(It can be simplified using $ XX \equiv \mathbb{1}$.)

\noindent\fbox{
\parbox{\textwidth}{\underline{Generalization}: For the case of $ n $ conjuctions, a similar constructions works with $ C^{\otimes n}X $ replacing Toffoli: \\ 
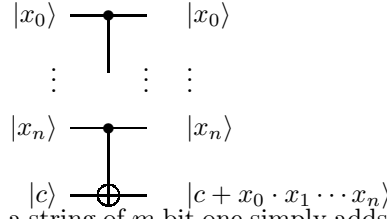
\centerline{
\Qcircuit @C=1em @R=2em {
\lstick{\ket{x_0}} & \ctrl{1} & \qw & \rstick{\ket{x_0}} \\
\lstick{\vdots} &  & \vdots & \rstick{\vdots} \\
\lstick{\ket{x_n}} & \ctrl{1}& \qw & \rstick{\ket{x_n}} \\
\lstick{\ket{c}} & \targ & \qw & \rstick{\ket{c + x_0 \cdot x_1 \cdots x_n}}
}
}
 
For Boolean functions whose output is a string of $ m $ bit one simply adds additional target qubits.

}
}
\subsection{Phase Oracles}
The same method can be used to construct the phase Oracle for $(-1)^{f(x)}$. This is done by replacing $C^{\otimes n}X$ by $C^{\otimes (n-1)}Z$ gates, for example the circuit for the phase gate of the function in Eq.~(\ref{e:f}) is shown in Fig.~\ref{fig:dcp}.
\begin{center}
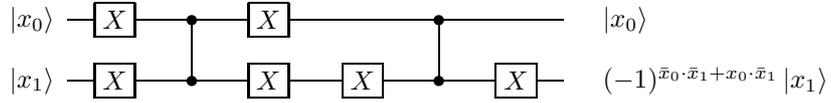
\begin{figure}[h]
    \centering
    \centerline{
\Qcircuit @C=1em @R=1em {
\lstick{\ket{x_0}} & \gate{X} & \qw & \ctrl{1} & \qw & \gate{X} & \qw & \qw & \qw & \ctrl{1} & \qw & \qw & \qw & \rstick{\ket{x_0}} \\
\lstick{\ket{x_1}} & \gate{X} & \qw & \ctrl{-1} & \qw & \gate{X} & \qw & \gate{X} & \qw & \ctrl{-1} & \qw & \gate{X} & \qw & \rstick{(-1)^{\bar{x}_0 \cdot \bar{x}_1 + x_0 \cdot \bar{x}_1}\ket{x_1}}
}
}
    \caption{The phase gate for the function in Eq.~(\ref{e:f}).}
    \label{fig:dcp}
\end{figure}
\end{center}
\section{The experiment: Additional details}\label{ap:error}
\subsection{Splitting the Grover Oracle}

Alice and Bob each get the same $n$-qubit Oracle. Alice uses the Oracle as is. Bob, uses the following algorithm to split the Oracle  to the two computers $B_1$ and $B_2$:

\begin{algorithm}[H] 
\KwResult{ The binary index of a desired element from the whole set. } 
\SetKwProg{BobGrover}{BobGrover}{}{} 
\BobGrover{$ (U_f) $}{
	$ U_{odd}, U_{even} \gets Divide(U_f) $ \;
	$ res_{1} \gets Grover(B_1,U_{even}) $ \; 
	$ res_{2} \gets Grover(B_2,U_{odd}) $ \; 
	\If { $ U_{even}(res_1) $ == 1 } {
		return $ res_1 $ + '0' \; 
	} 
	return $ res_2 $ + '1' \;
} 
\caption{Distributed Grover Algorithm} 
\end{algorithm}
{\subsection{A depolarizing channel model for the errors}\label{s:error}
}

A simple model of the noise in circuits is in terms of depolarizing channels\footnote{The condition on $\lambda$ guarantees complete positivity, see, \href{https://en.wikipedia.org/wiki/Quantum_depolarizing_channel}{Wikipedia}  }:
\begin{equation}
    \label{e:mixing}
    \rho \mapsto \lambda \rho + \frac{1-\lambda}{2} \mathbb{1}, \quad {-1/3\le \lambda\le 1}
\end{equation}
The probabilities that a qubit $\ket \psi$  gives the correct (1) /incorrect (0) answers are
\begin{equation}\label{e:lp}
    p_{0}=\frac{1-\lambda} 2\quad p_1=\frac{1+\lambda} 2, \quad 
\end{equation}
Assuming independence, the probability for $m$ bit-flip errors in $n$ qubits register is
\begin{equation}
 \binom{n}{m} p_0^mp_1^{n-m}
\end{equation}
The expected number of errors in the register is
\begin{equation}\label{e:np}
   n p_0
\end{equation}
and the probability for the correct answer in all $n$ qubits  is
\begin{equation}
\left(\frac{1+\lambda} 2\right)^n
\end{equation}
\subsection{Decoherence and gate errors}

The decoherence associated with each time step is represented by a polarizing channel with  $\lambda\mapsto \mu_c$ and the gates errors  gate by a polarizing channel with  $\lambda\mapsto \mu_g$. The composition of polarizing channels is ordinary multiplication.   It follows that for a circuit with $T$ time steps and $N_g$ noisy gates, is a channel where
\begin{equation}
    \lambda\approx \mu_c^T\mu_g^{N_g}
\end{equation}
Since the dominant error is in the 2-qubits gates,  $N_g$ is the number of 2-qubits gates.

From Table~\ref{t:ibm} 
\begin{equation}
  { \log \mu_c=-\frac R {T_c}\approx - 3\times 10^{-3} \Longrightarrow \mu_c\approx 0.997},\quad  \mu_g=1-\epsilon\approx 0.994
\end{equation}

{An estimate for $\lambda$  is then}\footnote{Since $T,N_g=O(1000)$ we need to keep 3  significant figures. }
\begin{equation}\label{e:ot}
    \lambda \approx (0.997)^{T} \cdot (0.994)^{N_g}
\end{equation}
This allows to compare the experimentally observed probability for bit flip error $p_0$ computed via Eq.~(\ref{e:np}) and the data in  Tables \ref{t:f4},\ref{t:error3} with the expected bit flip error $\bar p_0$ computed via  Eqs.~(\ref{e:lp}) and (\ref{e:ot}) using the IBM machine data in Table \ref{t:ibm}. The qualitative agreement between $p_0$ and $\bar p_0$ gives some support to the noise model as depolarizing channel.

\begin{table}[h]
\begin{center}
    \begin{tabular}{|l|c|c|c|}
    \hline
    Qubits    & $p_0$ & $\bar p_0$& $\lambda$ \\ 
    \hline
    4   & 0.50 & 0.47& 0.05\\
    3 (odd) & 0.30 &0.23&0.54\\
    \hline
    \end{tabular}
    \caption{The bit flip error $p_0 $ computed from the histogram and $\bar p_0 $ computed from the Machine data.}\label{t:3-4}
\end{center}
\end{table}

\end{appendices}

\bibliographystyle{plain}
\bibliography{ref.bib}

\end{document}